# The Laboratory of Mechanics and Acoustics in Marseilles (France) : from the first world war to the present day

S. Meunier, D. Habault, E. Friot, Ph. Lasaygues, H. Moulinec, C. Vergez
Aix Marseille Univ, CNRS, Centrale Marseille, LMA, 4 impasse Nicolas Tesla, CS 40006, 13453 Marseille Cedex 13, France

**Abstract**

The Laboratory of Mechanics and Acoustics in Marseilles (France) was created in 1941, under the name of *Centre de Recherches Scientifiques, Industrielles et Maritimes* (CRSIM). But it was actually issued from the French Naval Research Center created in Toulon by the French Navy to work on submarine detection during World War I. LMA is therefore the result of a long and quite amazing story with several moves and even more name changes. It benefited from all these events and is today established in a new campus with large facilities specially designed for its latest research activities. This article presents the story in some details, summarize the evolution of the research domains through all these years and finally gives a description of the LMA today.

## 1. Introduction

Today the Laboratory of Mechanics and Acoustics (LMA) is dedicated to Solid Mechanics and Acoustics. It is a research unit linked to three main organisms which are *Aix-Marseille University* (AMU), the National Center for Scientific Research (*Centre National de la Recherche Scientifique*, CNRS) and the *École Centrale Méditerranée* (ECM). Its staff includes about 130 people (researchers, engineers, technicians, administrative staff, doctoral students). Its early origin dates back to the beginning of the 20th century, but it is during the second world war that it became a CNRS research laboratory. Its story is detailed below following previous texts, mainly the writings of Claude Gazanhes ([1], [2]). Paragraph 2 presents the history of the lab. Paragraph 3 describes the evolution of the main themes from the beginning up to 2000. Paragraphs 4 and 5 presents the research activities and environment of the present lab.

## 2. History of the LMA

The very origins of the laboratory go back to the *Laboratoire de la Guerre Sous-Marine* (Submarine Warfare Laboratory) which was created in Toulon in 1917 by the French Navy. It must be noted that it is in this laboratory that Paul Langevin developed the first high-powered ultrasonic transmitters for the detection of enemy submarines.

At the end of the war, in 1920, the missions of this laboratory expanded and it became the *Laboratoire du Centre d'Etudes de Toulon* (Laboratory of the Toulon Study Center, LCET). Its scientific direction was entrusted to François Canac who played a major role in the history of the laboratory of which he was the founder and then the director until 1958.

At that time F. Canac was strongly inspired by the National Physical Laboratory, Teddington (UK). He decided that the LCET, far beyond its activities for the French Navy, should be a research unit committed to both theoretical research and technological survey, in order to optimize its capacity of developing, for any state department or industrial sector, applications or prototypes resulting from the latest discoveries, what is called today "technology transfer".

The first major date of this story is 1940. At the beginning of the Second World War, the laboratory was renamed *Centre de Recherches de la Marine* (CRM). At the end of June 1940, the CRM, staff and equipment, was hurriedly embarked on a ship to Oran and then a train to Algiers. It seemed that the CRM would disappear in the turmoil. But its director was an energetic man with a strong personality. In August 1940, F. Canac proposed that the laboratory be placed under the supervision of the Ministry of Public Instruction. He spoke in praise of the laboratory, its working methods, its scientist staff, its equipement, its ability to tackle high-level research as well as technical problems. And he was heard... At that time, the CNRS had just been created in order to bring together research units at a national level. Its administrative and financial situation was still under study but everything happened very quickly. On January 1, 1941, the CRM became the first laboratory of the CNRS out of Paris, with a new name *Centre de Recherches Scientifiques Industrielles et Maritimes* (Center for Scientific, Industrial and Maritime Research, CRSIM).

This lab can be seen as the ancestor of LMA. The research at CRSIM was divided into five major domains: Acoustics, Optics, Chemistry-corrosion, X-rays, Psychotechnics. The Acoustics part was strongly developed, notably under the influence of F. Canac who was passionately interested in the Acoustics of the Ancient Theaters [3]. It included electroacoustics, vibrations, architectural acoustics, ultra-sounds in air and water. The CRSIM already had many facilities: a large anechoic chamber (13 x 5 x 5 m3), a smaller one (4 x 2 x 2 m3), a reverberation room (10 x 2.6 x 2.6 m3), a ripple tank for the study of room acoustics, a 18m-channel with an absorbing tank. Figure 1 presents a view of the large anechoic chamber and reminds us of the difficulties of the time: The absorbing walls were made of pyramidal elements covered with blankets supplied by the Military Health Service.

*Figure 1: the anechoic chamber of the CRSIM*

Another important date is 1962. At that time, the lab moves to a new CNRS campus and it becomes the *Centre de Recherches Physiques* (Physical Research Center, CRP) with 4 departments: Mechanics Physics and Acoustics, Wave Visualization, Automation, Structures of crystalline bodies. Thanks to this operation, the laboratory expands and benefits from the installation of a new powerful equipment for acoustics: Among others, a large anechoic room for sound propagation, a smaller one for psychoacoustic tests (both with highly-absorbant materials), a station for architectural acoustics built according to international standards and including two adjacent reverberant rooms for studying the acoustics transparency, a music studio with a piano, several tanks dedicated to underwater acoustics and ultra-sound propagation.

In 1973, due to new arrivals, the laboratory refocuses on Solid Mechanics and Acoustics and becomes the LMA. This gives the opportunity to develop activities in Mechanics of continuous media, computer science as well as real-time techniques and to build facilities for mechanical experiments.

In 2012, the *Laboratoire de Contrôle Non Destructif* (Non-Destructive Control Laboratory, LCND-AMU), joins the LMA. Since then, the LMA has a branch and premises at the *Institut Universitaire de Tehcnologie* in Aix-en-Provence (a university mechanical engineering school).

The last date of the LMA story is 2015, when the LMA moves to a new campus, namely the Technopôle de Château-Gombert. This gives the opportunity to become closer to fluid mechanics and physics laboratories, as well as to the engineering schools *École Centrale Méditerranée* and *Polytech-*

*Marseille*. To strengthen and facilitate the administrative relations, in 2018, the LMA becomes a joint research unit under supervision of AMU, CNRS and ECM.

## 3. The evolution of research themes

Since its creation, the laboratory has been committed to developing research in both theory and applications. Here we summarize the development of the Acoustics theme and, more briefly, that of Solid Mechanics.

### 3.1. Acoustics

The laboratory report produced in 1944 [4] gives an idea of the studies developed in acoustics at that time. It includes work on sound protection against industrial and urban noise, room insulation (national broadcast radio studios, airplane cabins), sound propagation above grounds (using anechoic room experiments), ultrasonic detection of internal defects in metallic pieces, soundings to determine ice purity of glaciers, ultrasound applications for chemical and biological treatment of oil. From the end of the war onwards, the areas of expertise at LMA have been partly modified, according to the arrival/departure of researchers and with the lab moving to new campuses. Here are some examples.

At the outset, the 2 main laboratory themes were **applied acoustics** and **room acoustics**.

In **applied acoustics**, in addition to studies on acoustic insulation, the activities were focused at first on hall acoustics (music and theater), wall transparency, resonators (in tunnel and church) and, later, on subway noise. This work, which has enabled the lab to acquire experimental know-how, is much more limited today, as the LMA mostly interacts with industry on longer-term projects, notably within the framework of PhD theses. This activity has also enabled the CRP to take part in discussions on standards and regulations concerning the noise environment in France and abroad. In **room acoustics** the very first works were focused on the acoustics of ancient theaters (including the famous Orange theater), with optical experiments in tanks. Later the activity included the perceptual evaluation of room acoustics, but it came to a halt in the early 2000s.

With the passing years, other activities strongly developed:

**Underwater acoustics**, where work was initially devoted to wave propagation in marine environments. A great deal of research has been carried out in this field, in response to requests from the French Defense Agency (*Direction Générale de l'Armement* - DGA) for target detection (sonar, submarines, etc) and characterization of marine environments and seabeds. The addition of a large tank (20m-long) in 1987 was the starting point for new studies, in collaboration with the French Navy (DCN), on topics related to the propagation in the presence of bathycelerimetry, for example defining scaled experiments to reproduce the fluctuations due to internal waves.

In **psychoacoustics**, a team was set up around 1985, developing close links with American and German laboratories. Initial studies focused on the active mechanisms of attention, sound localization and loudness. With the arrival of young researchers, the activity was extended to the sound quality of urban and industrial noise (perceptual criteria of annoyance), the perceptual evaluation of sound reproduction systems, and more recently cochlear implants.

**Active noise control** benefited of the creation of a reinforced team in the 1980s. Numerous theoretical and experimental studies were carried out, mostly within the framework of European projects linked to transport noise; control also included real-time restitution of sound environments. In 1993, Alain Roure was awarded the CNRS crystal for his work in active control. The world's first patent for noise-canceling headphones was filed by LMA in 1985, the starting point for the Technofirst company.

**Musical acoustics** is also an activity that has grown considerably over the years, taking advantage in particular of new computer tools. It rapidly extended from the study of harmony and melody concepts to computer music, with sound analysis and synthesis, real-time control and the monitoring of sound and musical signals. Also included was the study of the physics of musical instruments (mainly wind instruments), in relation with the development of sound synthesis through physical modeling. The LMA was one of the first laboratories to develop the use of wavelets in the audible domain. In 1999, the CNRS awarded its gold medal to Jean-Claude Risset, senior researcher at the CNRS (the first Engineering Sciences researcher to receive this distinction).

**Acoustic and elastic wave propagation**, where the first studies concerned the propagation and diffraction of acoustic waves in air, then extended to solids and biological tissues, with direct problems (assessment of noise levels in outdoor environments, diffraction by obstacles, characterization of acoustic radiation from vibrating structures, non-linear absorbers, etc) and inverse problems (characterization of industrial objects and human tissues from acoustic measurements).

**3.2 Solid Mechanics**

Theoretical studies concerned mainly continuum mechanics and non-linear vibrations. They were divided into two areas: the first, entitled "Materials", covers the behavior of heterogeneous materials and structures, the resistance and monitoring of laminated composites, and the behavior of elastomers; the second, entitled "Structures", covered the study of contact, friction and interfaces, non-linear vibrations, and the multiphase mechanics of divided media. In 2004, Pierre Suquet, senior researcher at LMA, was elected a member of the French Academy of Sciences.

**4. The LMA today**

Today the activities of LMA are divided into 3 groups respectively named "Materials and Structures", "Waves and Imaging", and "Sounds".

The first one is concerned with all the activities in Solid mechanics, with theoretical, numerical and experimental aspects. Topics covered include nonlinear dynamics of structures, mechanics of heterogeneous media, interface mechanics (bonding) and contact/friction problems, multiphysics modeling and mechanical strength of laminated composite materials, and design of composite structures.

The research carried out by the Waves and Imaging team is aimed at developing acoustic methods for modeling, characterizing and/or imaging heterogeneous media, using both fundamental and applied approaches. The works focus on propagation in microstructured, porous and random media, nonlinear propagation, acoustics in marine environments, seismic imaging, non-destructive testing and evaluation, and biomedical ultrasound.

The Sounds team is concerned with all the aspects of surrounding audible sounds, which include their production, propagation, control, synthesis and perception. The works focus on the synthesis and control of acoustic fields, the study of the sound production by self-oscillating dynamic systems, the design and quality of musical instruments, and the mechanisms of auditory perception.

In addition, more recently, strong inter-team collaborations have been developed in non-linear dynamics (physical modeling of musical instruments, development of noise and vibration absorbers based on the concept of irreversible energy transfer), the use of ultrasound to stimulate the auditory nerve (with application to cochlear implants), and non-linear wave propagation in fluid and solid media (such as trumpets and NDT applications for concrete). An initiative has also been launched in connection with the deployment of bistable micro-architecture structures (propagation of state-transition fronts). Another cross-disciplinary theme is emerging in the field of instrument making, in relations with new materials and manufacturing processes (combining skills in the design of composite structures and the physics of musical instruments).

In 2015, moving in a new building has enabled the LMA to design experimental rooms closely suited to its acoustics research. This optimal equipment includes in particular a set of 3 anechoic large rooms : An anechoic room specially designed for the study of low-frequency active anechoicity (Figure 2), another one dedicated to psychoacoustic experiments, and a semi-anechoic room coupled, through the floor, to an excitation room, devoted to large-scale projects directly linked to industrial applications. This equipment is completed with a set of 3 audiometric booths, one of which is faradized (for electrophysiology measurements), a set of studios (for musical acoustics and art/science projects, Figure 3), a listening room for normalized restitution in non-anechoic conditions, and various large rooms for vibroacoustic experiments. A field-synthesis room enables the reproduction of very low-frequency sounds at perceptible levels.

*Figure 2: the large anechoic room for active control of low-frequency noise*

*Figure 3: music studio with 3D sound reproduction facility*

The laboratory is also equipped with scanners, various rooms for ultrasound experiments (room with anti-vibration slab for airborne ultrasound, laser/photo-acoustic room), research ultrasound scanners, optical microscopes and high-speed cameras. Work on biological materials, more recently developed at LMA, benefits of a special equipment (L2 cell-culture room).

Furthermore the facilities in mechanics include a mechanical testing room (88.5m²), a tribology and microscopic observation room (10m²), an X-ray microtomograph (shared with other laboratories on the campus) and facilities for designing, dimensioning, manufacturing and testing composite material structures.

## 5. The research environment

As for all labs, the environment of LMA includes cooperations with other research labs, industrial partners, teaching activities and knowledge dissemination.

The LMA has long-term collaborations with numerous mechanical and acoustic laboratories in France and abroad. As far as industrial partnerships are concerned, the LMA has followed in the footsteps of its founder, cultivating strong links with applied research, both with major groups and smaller companies. The main areas of collaboration have been and are still Transportation (Aeronautics and Space), Health, Environment, Energy and Music. Some recent examples include work in seismics/geoacoustics, musical acoustics (with Yamaha and Buffet-Crampon for instrument manufacturing), mechanics (with the deployment of flexible structures), partnerships on Nuclear Energy, and work on cochlear implants.

Directly linked to the university and engineering schools, the laboratory is also highly involved in training and teaching activities (mainly AMU and ECM but not only). The acoustics teams are responsible for an international Master's degree entitled WAVES (Waves, Acoustics, Vibrations, Engineering and Sound).

From the outset, the laboratory has also been involved in disseminating information. It is said that as early as 1948, for example, CRSIM organized a colloquium on computation methods for mechanical problems at which Burgers presented his equation for wave propagation in the non-linear regime. F. Canac was the first French editor of Acustica and one of the founders of the so-called GALF group which later became the *French Acoustical Society* (SFA). Since those days, many LMA members have been serving as active members of the Bureau and/or Groups of the SFA, and later EAA.

**Acknowledgments**
The recent photos were taken by Alain Rimeymeille, who also provided us with the old photo of the anechoic chamber, for which we would like to thank him warmly.

**A few external links**
on the CRSIM : *https://images.cnrs.fr/video/1267*
on the LMA 50$^{th}$ birthday *https://images.cnrs.fr/video/178*
the LMA web site *www.laboratoire-mecanique-acoustique.fr*